\documentclass[aps,pre,twocolumn,nofootinbib,longbibliography]{revtex4-1}
\usepackage{graphicx}
\usepackage{color}
\usepackage{epsfig}
\usepackage[caption=false]{subfig}

\begin{document}

\title{Non-Gaussianity of van Hove Function and Dynamic Heterogeneity Length Scale}
\author{Bhanu Prasad Bhowmik}
\email{bhanupb@tifrh.res.in}
\author{Indrajit Tah}
\email{indrajittah@tifrh.res.in}
\author{Smarajit Karmakar}
\email{smarajit@tifrh.res.in}
\affiliation{
Tata Institute of Fundamental Research, 
36/P, Gopanpally Village, Serilingampally Mandal,Ranga Reddy District, 
Hyderabad, 500107, Telangana, India }
\begin{abstract}
Non-Gaussian nature of the probability distribution of particles' displacements
in the supercooled temperature regime in glass-forming liquids are believed to
be one of the major hallmarks of glass transition. It is already been established
that this probability distribution which is also known as the van Hove 
function show universal exponential tail. The origin of such an exponential
tail in the distribution function is attributed to the hopping motion of 
particles observed in the supercooled regime. The non-Gaussian nature can also
be explained if one assumes that the system has heterogeneous dynamics in space
and time. Thus exponential tail is the manifestation of dynamic heterogeneity. 
In this work we directly show that non-Gaussanity of the distribution of 
particles' displacements occur over the dynamic heterogeneity length scale
and dynamical behaviour course grained over this length scale becomes homogeneous. 
We study the non-Gaussianity of van Hove function by systematically coarse 
graining at different length scale and extract the length scale of dynamic 
heterogeneity at which the shape of the van Hove function crosses over from 
non-Gaussian to Gaussian. The obtained dynamic heterogeneity scale is found 
to be in very good agreement with the scale obtained from other conventional 
methods.           
\end{abstract}
\maketitle

\section{Introduction}
\begin{figure}[!h]
\hskip -0.5cm
\includegraphics[scale = 0.35]{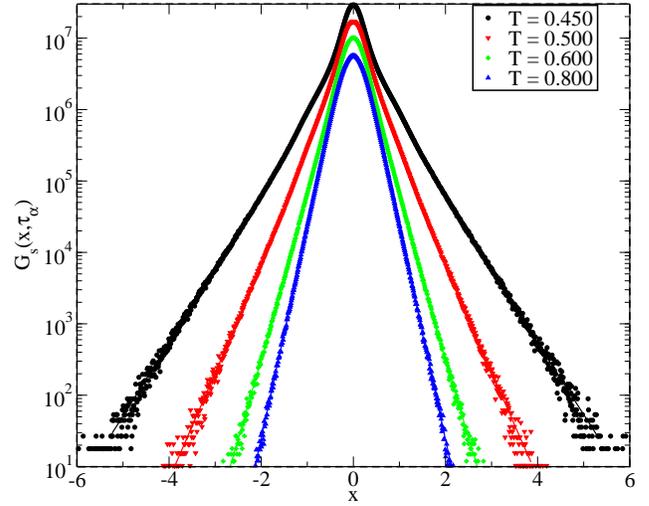}
\caption{van Hove function for 3dKA model for different temperatures. The 
non Gaussian tail becomes very prominent with decreasing temperature. The
line passing through the data points are the one obtained using 
Eq.\ref{lucyIteration} (see text for details).}
\label{vanHove3dKA}
\end{figure}

The distribution of displacement of particles, also known as 
van Hove function \cite{PhysRev.95.249} shows non-Gaussian behaviour in the supercooled
temperature (density) regime as demonstrated in Fig.\ref{vanHove3dKA}
for the well studied Kob-Andersen model of glass forming liquids. 
This has already been established as one of the main manifestations 
of dynamic heterogeneity in the dynamics of supercooled glass-forming 
liquids. Non-Gaussianity is also observed in many out of equilibrium 
systems that show glass like dynamical behaviour {\it e.g.}, vibrated 
granular medium. It has been observed that the non-Gaussianity in the 
van Hove function appears generically as an exponential tail \cite{07CBK}.  
A continuous time random walk (CTRW) approach to such an observation, 
suggests that one can understand the appearance of such an exponential tail
in the van Hove distribution, if one assumes that the particles in the 
systems are performing jump like (hopping) motion between two successive 
localized diffusive motions. The range of the exponential tail depends 
on the the diffusivity of the local motion and the waiting time 
distribution for the successive hopping motions. By carefully choosing 
the parameters one can fit the observed non-Gaussian feature of the van 
Hove function in a wide variety of systems\cite{07CBK}.

A different approach to rationalize the observed non-Gaussian behaviour in 
van Hove function is to assume spatial heterogeneity in the diffusion 
constants of constituent particles. Historically this idea has been 
introduced to understand the break down of Stokes-Einstein relation 
in supercooled liquids. In liquids, the Stokes-Einstein (SE) relation 
\cite{SER08HM,SER05Einstein,SER87LL} relates the shear viscosity ($\eta$)
to the translational diffusion constant ($D$) of a particle as 
$D = \frac{AT}{\eta}$, where $A$ is a constant. The value of this 
constant depends on the details of the particle and boundary conditions 
and $T$ is the temperature. This relation is derived for a 
probe particle in the hydrodynamic limit, but the SE relation is found
to hold for the self diffusion of liquid particles also at 
high temperatures \cite{Th93HS}. In the supercooled temperature regime the
relation is found to break down  \cite{SEB81Pollack,SEB90Rossler,SEB92FGSF,
SEB94KA,SEB95TK,SEB97CS,SEB00EdigerReview,SEB01MFCR,SEB01ML,SEB03SBME,
SEB06Chen,SEB06BPS,SEB09Xu,RevModPhys.83.587,SEB10M}. 
Initially Stillinger and Hodgdon \cite{Th93HS} and later Tarjus and 
Kivelson \cite{SEB95TK} phenomenologically proposed that by considering 
supercooled liquids to consist of mobile ``fluid-like'' and less mobile 
``solid-like'' regions, the break down of the Stokes-Einstein relation 
can be explained naturally as the average diffusion constant is 
predominantly determined by the ``fluid like'' regions whereas the 
average relaxation time is dominated by the ``solid-like'' 
regions. The clusters consisting of ``fluid-like'' or ``solid-like'' 
particles has been detected in many different studies \cite{DH-direct,DH-direct1,DH-direct2,1742-5468-2016-7-074003}.

For an example, if one considers a system with regions that have two 
diffusivities - one for ``solid like'' ($D_s$ ) and the other for 
``liquid like'' regions ($D_l$). The the distribution of diffusivity 
can be written as $p(D) = A\delta(D_s) + B\delta(D_l)$ where $A$ and 
$B$ are fixed by the normalization condition and the amount of solid 
like and liquid like regions. The van Hove function will then read as
\begin{equation}
G_s(x,t) = \int dD p(D) g(x|D),
\label{vanInt}
\end{equation}
where $g(x|D) = \frac{1}{\sqrt{4\pi D t}}\exp{\left(-\frac{x^2}
{4Dt}\right)}$ is the distribution of displacement of particles 
undergoing diffusive process. With Eq.\ref{vanInt} it can be shown 
that the van Hove function will have a long tail and depending on 
the distribution of the $p(D)$, the tail of the distribution can 
be either exponential or even Gaussian \cite{12WKBG}. In general 
the exponential tail has been reported \cite{07CBK} which, as 
emphasized in \cite{12WKBG}, might be due to the small range of 
the data. 

A natural question that arises in this context is as follows.
If one measures dynamical quantities like van 
Hove function using coarse-grained over certain length scale, will 
the dynamics looks spatially homogeneous? For example, if we 
calculate van Hove function coarse-grained over some specific 
length scale, will it loose its non Gaussian tail and will 
become Gaussian. If the answer is affirmative, then this will 
give us a natural procedure to extract the underlying dynamic 
heterogeneity length scale. This will also directly prove 
the picture of supercooled liquid being mosaic structure 
of fluid-like and solid-like regions with size of these structures
being equal to the dynamic heterogeneity length scale. 
In Ref.\cite{srikanthPRL2016}, it is shown that if one calculates 
the wave-vector dependent $\alpha$-relaxation time, 
$\tau_\alpha(q)$ in the supercooled temperature regime, then 
one finds that Stokes-Einstein relation does not break down 
above a characteristic wave vector, $q^*$ which depends on the 
studied temperature, $T$. The inverse of this characteristic 
wave vector defines a length scale, $\xi^*(T) = 2\pi/q^*(T)$. 
This length scale, $\xi^*(T)$ is found to be same as that of 
the dynamic heterogeneity length scale obtained from the analysis
of four-point dynamic susceptibility \cite{chandan92}, $\chi_4(T,t)$ calculated at 
$\alpha$-relaxation time, $t=\tau_\alpha(q=q_0)$. $q_0$ is the 
position of the first peak in the static structure factor, $S(q)$.
This study also suggests that dynamics coarse-grained over 
dynamic heterogeneity length scale might look homogeneous, leading
to a direct measure of dynamic heterogeneity length scale from 
experimental data. 

The goal of the present work is to measure the non-Gaussian behaviour 
of van Hove function by systematically coarse graining the dynamics
over different length scale to study the cross over from non-Gaussian 
to Gaussian form  at some characteristic length scale. Then understand 
the relation between this characteristic length scale with the dynamic 
heterogeneity length scale obtained from the conventional methods. 
For systematic coarse-graining, we have employed the block analysis 
\cite{indrajitPRL2017} which has recently been used very successfully 
to perform finite size scaling analysis of four-point susceptibility, 
$\chi_4(t)$.  

The rest of the paper is organized as follows. First we will discuss
the model glass-forming systems that are studied in this work and 
the details of the simulation performed. Then we will discuss the 
correlation functions and the method of block analysis that has been 
employed to do the systematic coarse-graining of the 
dynamics. Next we will discuss how distribution of diffusion constants
are extracted from the van Hove function using iterative methods. 
Finally, we will discuss the results and conclude with possible 
application of this results for experimentally relevant systems.  

\section{Models and Simulation Details}
We have studied three different model glass forming liquids in three 
dimensions for $N = 108000$. The model details are given below:
\vskip +0.3cm
\noindent{\textbf{3dKA:}} The model glass former, we have studied is 
the Kob-Anderson $80:20$ \cite{KA} Lenard-Jones Binary mixture. 
This model was first introduced by Kob-Anderson to simulate 
$Ni_{80}P{20}$. This model has been studied extensively by many people 
and found to be a very good glass former in three dimensions. 
The interaction potential is given by
\[ V_{\alpha\beta}(r) = 4.0\epsilon_{\alpha\beta}[(\frac{\sigma_{\alpha\beta}}{r})^{12} - (\frac{\sigma_{\alpha\beta}}{r})^6]\]
where $\alpha,\beta \in \{ A,B \}$ and $\epsilon_{AA} = 1.0$, 
$\epsilon_{AB} = 1.5$, $\epsilon_{BB} = 0.5$; $\sigma_{AA} = 1.0$, 
$\sigma_{AB} = 0.80$,   $\sigma_{BB} = 0.88$. The interaction potential 
is cut off at $2.50\sigma_{\alpha\beta}$ and the number density is 
$\rho = 1.20$. Length, energy and time scale are measured in units 
of $\sigma_{AA}, \epsilon_{AA}$ and $\sqrt{\frac{\sigma_{AA}^2}
{\epsilon_{AA}}}$. For Argon these units corresponds to a length 
of $3.4 \AA$, an energy of $120Kk_{\beta}$ and time of  
$3\times 10^{-13} s$. We have done simulation in the temperature 
range $T \in \{ 0.430, 0.800 \}$.

\noindent{\textbf{3dIPL:}} In this model, the inter particle interaction 
potential is modeled as purely repulsive inverse Power Law form. This model 
has been studied in \cite{10PSDPRL}. We have studied in the temperature range $T \in \{ 0.450, 0.800 \}$
The interaction potential is given by 
\[ V_{\alpha\beta}(r) = 1.945\epsilon_{\alpha\beta}[(\frac{\sigma_{\alpha\beta}}{r})^{15.48}]\]
All the parameters and interaction cut-off is same as the 3dKA model.

\noindent{\textbf{3dR10:}} This is a \textbf{50:50} binary mixture \cite{2dR10} 
interacting via the pair wise interaction potential
\[ V_{\alpha\beta}(r) = \epsilon_{\alpha\beta}[(\frac{\sigma_{\alpha\beta}}{r})^{10}]\]
Here $\epsilon_{\alpha\beta} = 1.0$, $\sigma_{AA} = 1.0$, $\sigma_{AB} = 1.22$,   
$\sigma_{BB} = 1.40$. The interaction potential is cut-off at 
$1.38\sigma_{\alpha\beta}$. The number density of particle is 0.85 and the 
temperature range studied is $T \in \{ 0.520, 0.800 \}$.

We use the modified leap-frog algorithm with the Berendsen thermostat to 
keep the temperature constant in the simulation runs. Any other thermostat 
does not change the results qualitatively as we are mostly interested in 
configurational changes in the system instead of momentum correlations. 
The integration time steps used is $dt = 0.005$ in the studied temperature 
range.

To characterize the dynamics, we have calculated two point correlation 
function $Q(t)$, which gives the amount of overlap between two configurations 
which are separated by time $t$.
\begin{equation}
Q(t) = \sum_{i=1}^{N}w(\left| r_i(0) - r_i(t)\right|),
\label{qt}
\end{equation}
where the window function $w(x) = 1$ when $x\leq a$ and $0$ otherwise. 
We choose $a = 0.3$ which is close to the plateau value of 
the mean square displacement. This parameter is chosen to remove possible
decorrelation that can happen due to vibrational motion of the particles
inside the cage formed by their neighbours. A different choice of
this parameter does not change the temperature dependence of the 
$\alpha$-relaxation time, $\tau_{\alpha}$. $\tau_{\alpha}$ is defined 
from the decay of $Q(t)$ as $\langle Q(\tau_{\alpha})\rangle = 1/e$. 
$\langle\ldots\rangle$ refers to ensemble average and averaging over 
different time origin. The fluctuation or variance of the overlap 
function $Q(t)$ is defined as four point susceptibility \cite{chandan92}.
\[ \chi_4^p(t) = \frac{1}{N}[\langle Q^2(t)\rangle] - \langle Q(t) \rangle^2] \]
Dynamic length-scale, $\xi_D$ can be obtained from the finite size 
scaling of peak height of $\chi_4(t)$ very reliably \cite{PNASUSA2009}. In this study, we
have taken the results of $\xi_D$ from Ref.\cite{indrajitPRL2017}. It is important 
to note that peak of $\chi_4(t)$ appears at $t=\tau_4$, which is very
close to $\tau_\alpha$. This also suggests that heterogeneity is maximum
at time scale close to the $\alpha$-relaxation time. In this study thus 
we will look at the van Hove function at the same time scale.

\section{Results}
We start with the formal definition of the van Hove correlation function 
\begin{equation}
G_s(x,\tau) = \left<\delta\left[ x - (x_i(\tau) - x_i(0)) \right]\right>,
\end{equation}
where the $\langle\ldots\rangle$ implies the averaging over the time origin
and different statistically independent samples. 
\begin{figure}[!htpb]
\hskip -0.2cm\includegraphics[scale = 0.37]{{vanHove3dKATauAlpha}.eps}
\vskip +0.2cm
\includegraphics[scale = 0.37]{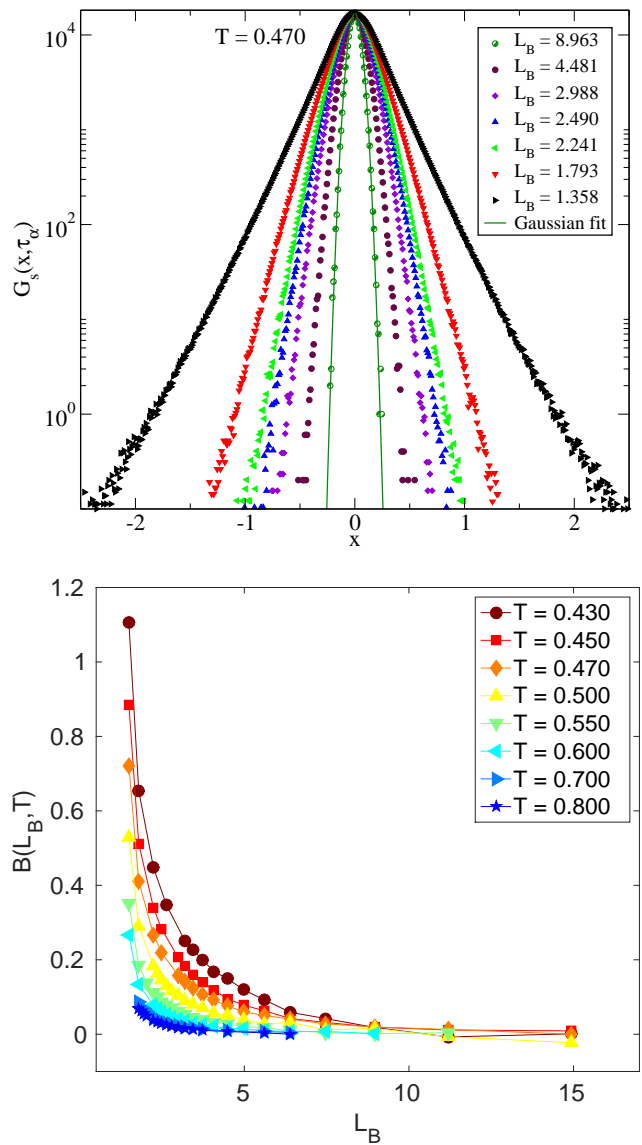}
\caption{{\bf Top Panel:} van Hove correlation function as a function of 
displacement for different coarse-graining block length, $L_B$. These
data are for 3dKA model. One can clearly see that as one increases the block length, the nature of the distribution
changes to Gaussian. All the curves are shifted to match the peak height.
The line passing through the data for $L_B = 8.963$ is a Gaussian fit 
to the data.  
{\bf Bottom Panel:} The Binder cumulant as a function of block length for 
different temperature. Binder cumulant becomes 0 for Gaussian 
distribution. Thus one can clearly see that for smaller block length
the distribution is very non Gaussian and becomes Gaussian at larger 
block length. The cross over length increases with decreasing temperature.}
\label{vanHoveFun}
\end{figure}
To perform systematic spatial coarse-graining of the dynamics, we have used 
the method of block analysis \cite{zippelius, indrajitPRL2017}. 
In this method, the whole simulation box is 
divided into smaller blocks of length, $L_B$. Thus with block size of $L_B$,
there will be $N_B = (L/L_B)^d$, number of blocks in the system. $L$ is the 
linear size of the simulation box and $d$ is the number of spatial dimensions.
This method is shown to be very attractive for doing finite size scaling of
four-point susceptibility $\chi_4(t)$ for extracting dynamic heterogeneity
length scale, $\xi_D$. Due to its simplicity, the method will be very
useful to study dynamic heterogeneity in experiments with colloidal particles.
In this work, we have defined a coarse-grained displacement as
\begin{equation}
\Delta x_j^B(\tau) = \sum_{i=1}^{n_j}\left[x_i(\tau)-x_i(0)\right],
\end{equation}
where $n_j$ is the number of particles in the $j^{th}$ block. Note that 
this number will be different for different blocks. Then we define the 
{\it blocked} van Hove function as
\begin{equation}
G^B_s(x,\tau) = \left<\sum_{j=1}^{N_B}\delta\left[ x -  \Delta x_j^B(\tau)\right]\right>,
\end{equation}
By varying the block length, $L_B$ we have studied how the non-Gaussianity 
changes with increasing block length. One thing to note is that, as one
increases the block length the total displacement decreases, this is easy
to understand as there is no center of mass displacement during the 
simulation and if we choose $L_B = L$, then the coarse-grained displacement
will be zero. As we are interested in the shape of the van Hove function 
this issue will not affect the analysis. 

In top panel Fig.\ref{vanHoveFun}, 
we have shown the van Hove function for $T = 0.470$ for 3dKA model for various 
coarse-graining block length, $L_B$. One can clearly see that with increasing
block length, the distribution becomes more and more Gaussian and for 
block length $L_B = 7.469$, at this particular temperature, the distribution
becomes completely Gaussian as shown by the fitted line to a Gaussian 
function. In the bottom panel, we have calculated the binder cumulant of the
distribution to measure the departure from the Gaussian form. The binder
cumulant is defined as
\begin{equation}
B(L_B,T) = 1 - \frac{\langle x^4\rangle}{3\langle x^2\rangle^2}
\end{equation}
which is zero for a Gaussian distribution. The average is done over 
the distribution, $G_s^B(x,\tau)$ for the respective block length 
$L_B$. The bottom panel of Fig.\ref{vanHoveFun}, shows that 
binder cumulant is non-zero for smaller block sizes and tends to become 
zero for larger block lengths. The approach to zero happens at larger 
block length for decreasing temperature.

Next we discuss how the underlying distribution of diffusion constants 
changes with coarse-graining volume. Before going in to the results, 
we explain briefly the method used to extract the distribution of 
diffusivity directly from the van Hove correlation function 
$G_s(x,\tau_{\alpha})$ using the Iterative algorithm suggested in 
Ref.\cite{74Lucy} and recently used in \cite{12WKBG} for the diffusion 
processes in biological systems and in \cite{skJCP2014, 1742-5468-2016-7-074003} 
for dynamics in supercooled liquids. If one assumes that particle 
displacements are due to diffusion processes and there is a distribution of 
local diffusivity $p(D)$, then formally we have
\begin{equation}
G_s(x,\tau_{\alpha}) = \int_0^{D_0} p(D). g(x|D). dD,
\label{vanHoveEq}
\end{equation}
where $g(x|D) = 1/\sqrt{4\pi D\tau_{\alpha}} \exp\left( -x^2/4D
\tau_{\alpha} \right)$ and $D_0$ is the upper limit of diffusion 
constant and will be equal to diffusivity for a free particle 
diffusion. 
\begin{figure}[!t]
\includegraphics[scale = 0.35]{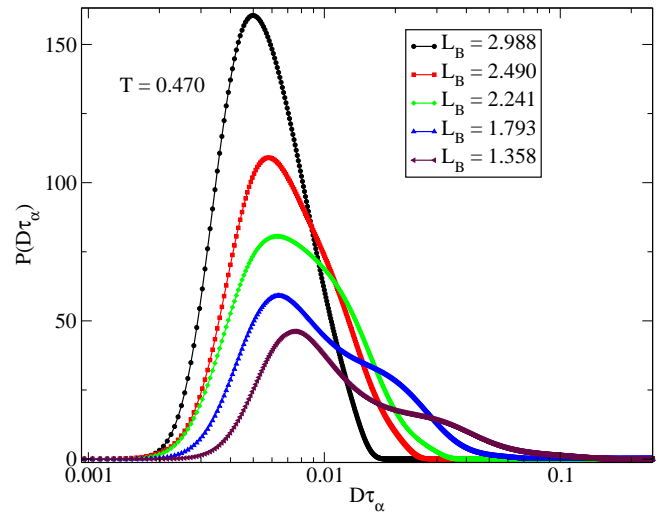}
\caption{Distribution of diffusion constants obtained for coarse-grained
van Hove function for the data in Fig.\ref{vanHoveFun}. One can clearly
see that distribution becomes bimodal to unimodal with increasing block
size.}
\label{distDiff}
\end{figure}
\begin{figure}[!b]
\includegraphics[scale = 0.34]{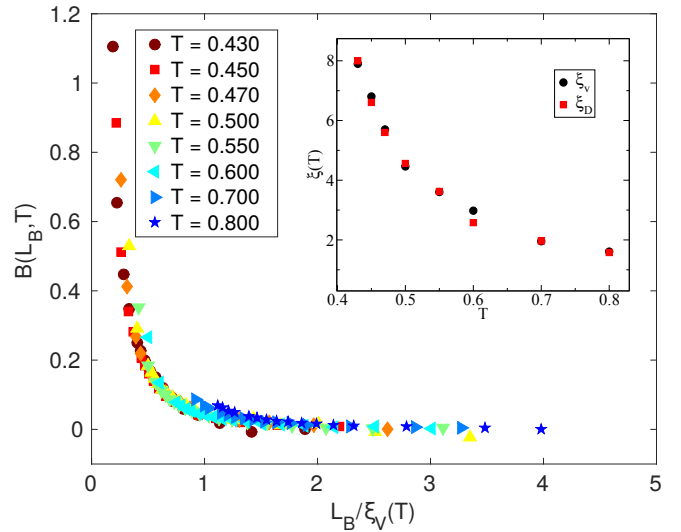}
\caption{Top panel: Finite size scaling of binder cumulant of the van Hove 
functions calculated for different block lengths. The scaling 
collapse is quite good. This suggests the existence of a cross
over length scale that grows with decreasing temperature.
Inset: Comparison of the cross over length scale, $\xi$
with dynamic heterogeneity length scale, $\xi_D$.}
\label{bcScaling}
\end{figure}
\begin{figure*}[htpb]
\includegraphics[scale = 0.87]{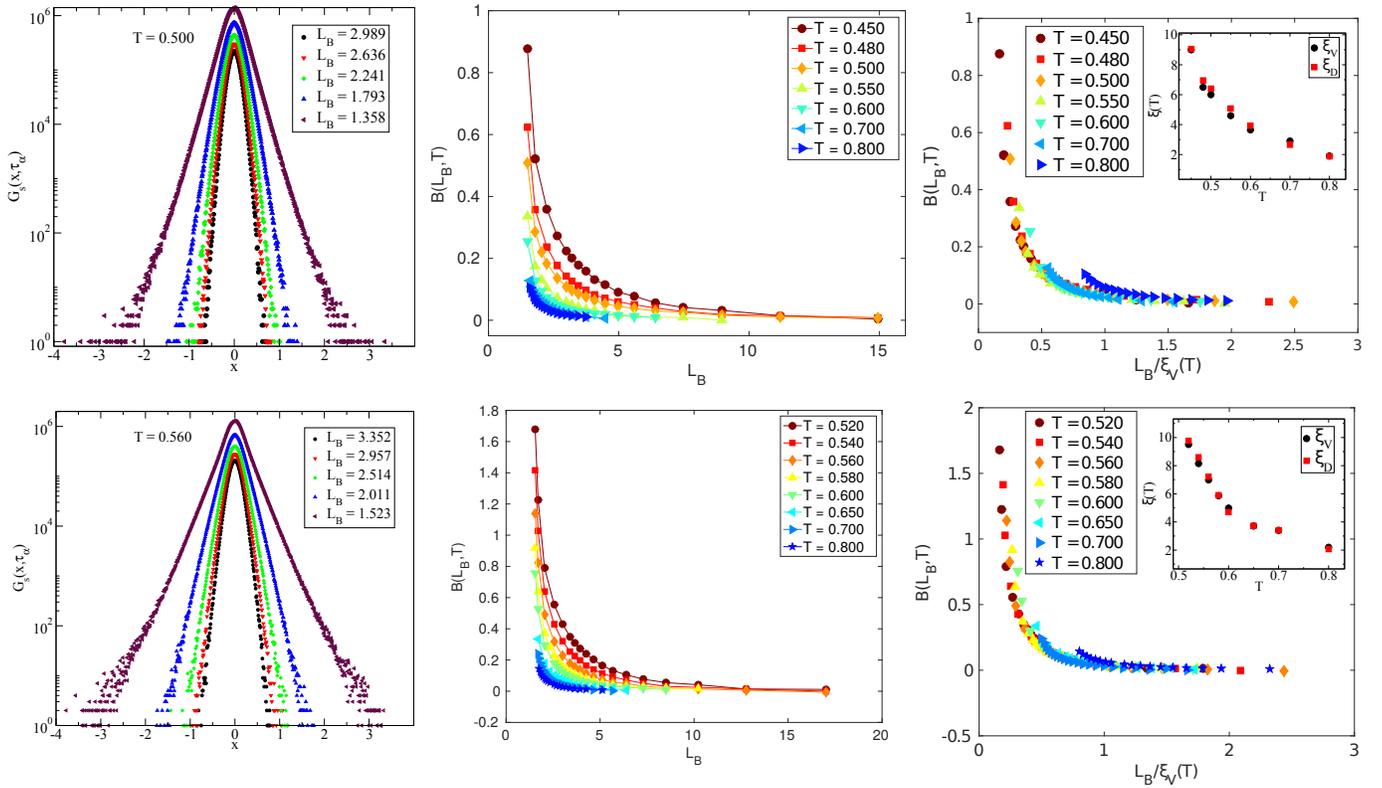}
\caption{Top left panel: Blocked van Hove function for different
block lengths for 3dIPL model. One can clearly see the shape 
of the van Hove function changes from non Gaussian with exponential 
tail to Gaussian. The curves are shifted for better clarity. 
Top middle panel: Binder cumulant of the blocked van Hove function
for 3dIPL model again quantitatively showing the non-Gaussian 
to Gaussian cross over with increasing block size. Top right panel:
Finite size scaling of binder cumulant of the van Hove 
functions. The scaling collapse is quite good. The inset shows 
the comparison of the cross over length scale, $\xi$
with dynamic heterogeneity length scale, $\xi_D$. Bottom panels:
Similar analysis done for 3dR10 model.}
\label{otherModels}
\end{figure*}
Now given the $G_s(x,\tau_{\alpha})$, $p(D)$ can be calculated 
using Lucy's Iterative method \cite{74Lucy} as 
\begin{equation}
p^{n+1}(D) = p^{n}(D)\int_{-\infty}^{\infty} \frac{G_s(x,\tau_{\alpha})}
{G^n_s(x,\tau_{\alpha})}g(x|D) dx,
\end{equation}
where $p^n(D)$ is the estimate of $p(D)$ in the $n^{th}$ iteration 
with $p^0(D) = (1/D_{avg})\exp(-D/D_{avg})$ is the initial input
guess distribution. Note that actual form of this guess distribution
does not change the final outcome. After $n^{th}$ iteration, 
$G^n_s(x,\tau_{\alpha})$ can be written as 
\begin{equation}
G^n_s(x,\tau_{\alpha}) = \int_0^{D_0} p^n(D). g(x|D). dD.
\label{lucyIteration}
\end{equation}
Similarly 
\begin{equation}
P^{n+1}(D\tau_{\alpha}) = P^{n}(D\tau_{\alpha})\int_{-\infty}^{\infty} 
\frac{G_s(x,\tau_{\alpha})}{G^n_s(x,\tau_{\alpha})}g(x|D) dx,
\end{equation}
where $p(D)dD = P(D\tau_{\alpha})d(D\tau_{\alpha})$. The choice of 
$D\tau_{\alpha}$ as our variable is due to the fact that $D$ changes by 
orders of magnitude in the studied temperature range whereas 
$D \tau_{\alpha}$ should change relatively modestly with 
decreasing temperature and it will be easier to compare the 
distribution obtained for different temperatures. 
In Fig.\ref{distDiff}, we have shown that obtained distribution of 
diffusion constants for the 3dKA model system at $T = 0.470$. The
distributions are obtained for the blocked van Hove function shown 
in Fig.\ref{vanHoveFun}. These results also clearly show that the
underlying distribution of diffusion constant becomes unimodal from
bimodal with increasing block length. 

We next perform the finite size scaling analysis of the binder cumulant
to obtain the coarse-graining length scale above which the van Hove 
function becomes Gaussian. We assume the following form of the scaling
function 
\begin{equation}
B(L_B,T) = \mathcal{G}[L_B/\xi_V(T)]. 
\end{equation}
In Fig.\ref{bcScaling}, we have shown the finite size scaling of the 
binder cumulant of the van Hove function calculated for different 
block sizes for the 3dKA model. The scaling collapse observed to be 
quite good. We then compared the obtained cross over length, $\xi_V$ 
with that of dynamic heterogeneity length scale, $\xi_D$ obtained 
from the block analysis of peak height of $\chi_4$. The $\xi_D$ data 
is taken from Ref.\cite{indrajitPRL2017}. The agreement between the two 
length scales over the whole temperature range suggest that the 
characteristic coarse-graining length is indeed same as that of 
the dynamic heterogeneity length scale.

To test whether the observations are generic for other glass forming 
liquids, we have performed similar analysis for two other model 
glass-formers, {\it e.g.} 3dIPL and 3dR10 model. In Fig.\ref{otherModels}, 
we have shown the results for 3dIPL and 3dR10 models. In top left panel 
of Fig.\ref{otherModels}, the van Hove function is plotted for 
$T = 0.500$ for different coarse-graining block length for 3dIPL 
model. For this model also one can see that the van Hove function
becomes Gaussian with increasing block length. In the top middle panel 
shows the binder cumulant of the blocked van Hove function. This also
clearly show that binder cumulant goes to zero with increasing block size
and the cross over happens at larger block sizes with decreasing 
temperature. The top right panel shows the scaling analysis of the 
data shown in the top middle panel. The inset in that figure shows the
comparison of the cross over length scale, $\xi_V$ with the dynamic
heterogeneity length scale, $\xi_D$ obtained again from finite size
scaling of four-point susceptibility. The data is taken from 
Ref.\cite{indrajitPRL2017}. In the bottom panels of Fig.\ref{otherModels},
we have shown similar analysis done for 3dR10 model. For both the models, 
one clearly sees that the cross over length scales are in very good agreement
with the dynamic heterogeneity length scales for the respective models.

To conclude, we have shown that the non-Gaussian nature of the van Hove
function can indeed be understood using the scenario of dynamic heterogeneity
manifested itself as regions of ``slow'' and ``fast'' moving particles over 
the characteristic relaxation time scale of the system. This study clearly
show that dynamical properties coarse-grained over heterogeneity length 
scale becomes homogeneous in complete agreement with the recent findings
\cite{srikanthPRL2016} where it was shown that wave vector dependent 
relaxation time obeys Stokes-Einstein relation below some characteristic 
wave vector which is inversely related to the dynamic heterogeneity 
length scale. Finally we show that the dynamic heterogeneity length scale
can be easily obtained by performing a careful finite size scaling of the 
binder cumulant calculated from the blocked van Hove function. We also 
show that the results obtained for one model system are generically 
applicable for many other glass-forming liquids also. We believe that 
this method of obtaining the heterogeneity length scale from van Hove 
function by systematically coarse-graining the dynamics will 
be very useful for analyzing data in experiments with colloidal particles. 
This method can also be used to study the growth of dynamic heterogeneity 
length scale for inter-facial water molecules near protein surface, cell 
membranes as these molecules are also shown to show heterogeneous dynamics.  

We thank Shiladitya Sengupta and Pinaki Chaudhuri for 
many useful discussion. We also thank Ananya Debnath for suggesting the
usefulness of this method for the study of dynamic heterogeneity of 
inter-facial water molecules near protein surfaces. Discussion with 
Srikanth Sastry is also acknowledged.

\bibliography{vanHoveDH} 
\bibliographystyle{apsrev4-1}
\end{document}